\magnification\magstep 1
\parskip 5pt plus 1pt minus 0.5pt
\def\init{\tabskip 0pt}
\def\crr{\cr\noalign{\hrule}}
\def\w{\widetilde}
\def\euler{\gamma_{\scriptscriptstyle E}}
\def\frc#1#2{{#1\over #2}}
\def\im{\mathop{\rm Im}}
\def\re{\mathop{\rm Re}}
\def\Disc{\mathop{\rm Disc}}
\def\det{\mathop{\rm det}}
\def\habs{{\vert h\vert}}
\def\min{{\rm min}}
\def\max{{\rm max}}
\def\EMA{{(\rm EMA)}}
\def\d{{\rm d}}
\def\e{{\bf e}}
\def\p{{\bf p}}
\def\r{{\bf r}}
\def\x{{\bf x}}
\def\y{{\bf y}}
\def\z{{\bf z}}
\def\0{{\bf 0}}
\def\D{{\bf D}}
\def\M{{\bf M}}
\def\N{{\bf N}}
\def\1{{\bf 1}}
\def\I{{\cal I}}
\def\L{{\cal L}}
\def\R{{\cal R}}
\def\V{{\cal V}}
\def\E{{\cal E}}
\centerline {\bf DIELECTRIC RESONANCES}
\medskip
\centerline {\bf OF LATTICE ANIMALS AND OTHER FRACTAL CLUSTERS}
\vskip 30pt
\centerline {by J.P. Clerc$^{(1,3)}$, G. Giraud$^{(1,3)}$,
J.M. Luck\footnote{*}{Corresponding author.
e-mail: luck@amoco.saclay.cea.fr}$^{(2,3)}$, and Th. Robin$^{(3)}$}
\vskip 30pt
\noindent (1) IUSTI, Universit\'e de Provence, Centre de Saint-J\'er\^ome,
avenue Escadrille Normandie-Niemen, 13397 Marseille cedex 20, France.
\smallskip
\noindent (2) C.E.A. Saclay, Service de Physique Th\'eorique,
91191 Gif-sur-Yvette cedex, France.
\smallskip
\noindent (3) X-RS, Parc-Club, 28, rue Jean-Rostand, 91893 Orsay cedex, France.
\vskip 50pt
\noindent{\bf Abstract.}
Electrical and optical properties of binary inhomogeneous media
are currently modelled by a random network
of metallic bonds (conductance $\sigma_0$, concentration $p$)
and dielectric bonds (conductance $\sigma_1$, concentration $1-p$).
The macroscopic conductivity of this model
is analytic in the complex plane of the dimensionless
ratio $h=\sigma_1/\sigma_0$ of the conductances of both phases,
cut along the negative real axis.
This cut originates in the accumulation of the resonances
of clusters with any size and shape.
We demonstrate that the dielectric response of an isolated cluster,
or a finite set of clusters,
is characterised by a finite spectrum of resonances,
occurring at well-defined negative real values of $h$,
and we define the cross-section which gives a measure of the strength
of each resonance.
These resonances show up as narrow peaks with Lorentzian line shapes,
e.g. in the weak-dissipation regime of the $RL-C$ model.
The resonance frequencies and the corresponding cross-sections
only depend on the underlying lattice,
on the geometry of the clusters, and on their relative positions.
Our approach allows an exact determination of these characteristics.
It is applied to several examples of clusters drawn on the square lattice.
Scaling laws are derived analytically, and checked numerically,
for the resonance spectra of linear clusters,
of lattice animals, and of several examples of self-similar fractals.
\vfill
\noindent To appear in the Journal of Physics A {\bf 29} (1996).
\hfill SPhT/96/042

\eject
\noindent {\bf 1. INTRODUCTION}
\smallskip
Frequency-dependent electrical and optical properties of inhomogeneous media
are often modelled by networks of random complex impedances.
The case of binary composite media has attracted a lot of attention
(see ref. [1] for a review).
Each bond of a regular lattice is independently attributed a random complex,
frequency-dependent conductance (or admittance) according to the binary law
$$
\sigma_{\x,\y}=\left\{\matrix{
\sigma_0&\hbox{with prob.}&p,\hfill\cr
\sigma_1&\hbox{with prob.}&1-p.\hfill\cr
}\right.
\eqno(1.1)
$$
The dimensionless complex ratio of the conductances of both phases,
$$
h=\frc{\sigma_1}{\sigma_0}
,\eqno(1.2)
$$
and the concentration $p$ are the essential parameters of the model.

The $\habs\ll 1$ regime describes several situations of interest.
As far as static (DC) properties are concerned, the $h=0$ limit embraces
the two well-known cases of the conductor-insulator mixture $(\sigma_1=0)$,
and the superconductor-conductor mixture $(\sigma_0=\infty)$.
In both limiting situations the macroscopic conductivity $\Sigma$
exhibits critical behaviour around the percolation threshold $p_c$,
$$
\matrix{
\Sigma(\sigma_1=0)\sim\sigma_0(p-p_c)^t\hfill&(p\to p_c^+),\cr
\Sigma(\sigma_0=\infty)\sim\sigma_1(p_c-p)^{-s}\hfill&(p\to p_c^-),
}
\eqno(1.3)
$$
where $s$ and $t$ are universal critical exponents.

Frequency-dependent (AC) electrical and optical properties
of metal-dielectric mixtures have attracted much interest recently.
Two models have been mostly investigated,
the $R-C$ model and the $RL-C$ model [1].
The dielectric component is modelled by perfect capacitors,
while the metallic bonds consist of a resistance in the first case,
of an inductance in series with a resistance in the second case.
The $RL-C$ model has been proposed to describe optical properties
of several kinds of inhomogeneous materials, such as cermets [2].
These materials undergo an optical transition at some value $p^*>p_c$
of the metallic fraction, where their dielectric constant changes over from
inductive to capacitive.
In both $R-C$ and $RL-C$ models, the frequency dependence of the ratio $h$
is such that the low-frequency regime corresponds to $\habs\ll 1$
(see section 2.2).

In the critical region, namely for small $h$
and near the percolation threshold,
the conductivity obeys a universal scaling law of the form
$$
\Sigma\approx\sigma_0\vert p-p_c\vert^t
\,\Phi_\pm\!\left(h\vert p-p_c\vert^{-(s+t)}\right)
\qquad(\habs\ll 1,\;\vert p-p_c\vert\ll 1)
,\eqno(1.4)
$$
where $\pm$ refers to the sign of $(p-p_c)$.

The present work aims at
a better knowledge of the analytic structure of the conductivity
in the complex $h$-plane, for any fixed concentration $p$.
In spite of its apparent academic character,
this point is essential for a quantitative understanding
of both the transient response
of the $R-C$ model to a time-dependent excitation,
and the resonance spectrum of the $RL-C$ model [1].
Hereafter we shall mostly focus our attention on the second situation.

Consider first a finite network, defined by putting on the bonds of a graph
conductances which assume two values, $\sigma_0$ or $\sigma_1$.
The conductance of this network between any two nodes
assumes the form $Y=\sigma_0 F(h)$, where $F(h)=N(h)/D(h)$ is a rational
function of the complex ratio $h$, i.e., $N(h)$ and $D(h)$ are polynomials,
whose degrees are roughly equal to the number of bonds of the network.
Furthermore the zeros of the conductance (i.e., the zeros of $N$)
and its poles (i.e., the zeros of $D$) are negative real numbers,
and they alternate, namely there is exactly one zero
between any two consecutive poles, and vice-versa.
These properties are at the origin of various
analytic representations and rigorous inequalities for the conductivity and
the dielectric constant of random media.
This body of knowledge is currently
referred to as the Bergman-Milton theory [3].

As a consequence, the singularities of the
macroscopic conductivity take place for $h$ real negative,
so that $\Sigma$ is analytic in the complex $h$-plane
cut along the negative real axis.
We introduce the notation
$$
\Disc\Sigma(h)=\frc{1}{i}\Bigl(\Sigma(h+i0)-\Sigma(h-i0)\Bigr)=2\im\Sigma(h+i0)
\eqno(1.5)
$$
for the discontinuity of the conductivity along this cut.

It is interesting to first look at the prediction of
the effective-medium approximation (EMA).
This old and very commonly used approximate scheme [4--6]
amounts to resumming the one-impurity
effects in a self-consistent way.
In the case of the square lattice the EMA formula reads
$$
\Sigma^\EMA=\sigma_0\Bigl((p-1/2)(1-h)+\sqrt{(p-1/2)^2(1-h)^2+h}\Bigr)
.\eqno(1.6)
$$
This expression shows that the conductivity is cut along a finite interval
$[h_\min, h_\max]$, with
$$
h_\min,\,h_\max=-\frc{1+4p(1-p)\pm 4\sqrt{p(1-p)}}{(2p-1)^2}
,\eqno(1.7)
$$
and that its discontinuity reads
$$
\Disc\Sigma^\EMA=2\sigma_0\bigl\vert p-1/2\bigr\vert
\sqrt{(h_\max-h)(h-h_\min)}\qquad(h_\min<h<h_\max)
.\eqno(1.8)
$$

It has been argued [1] that the exact conductivity of the random binary model
has a cut with a non-vanishing discontinuity
along the whole negative real axis.
The bulk of the discontinuity is expected to be roughly given by eq. (1.8),
while its tails are expected to be much weaker,
and to become exponentially small
as either $h\to 0^-$ or $h\to -\infty$, in analogy with the Lifshitz tails
of the density of states of electrons and phonons in disordered solids [7].
The following law for $D$-dimensional lattice problems
has been proposed in ref. [1]:
$$
\Disc\Sigma\sim\exp\left(-C\habs^{-D/2}\right)\qquad(h\to 0^-)
,\eqno(1.9)
$$
while a more recent investigation [8]
rather suggests the above law with formally $D=1$,
independently of the dimension $D$.

We propose to get a novel kind of physical insight into
the cut of the conductivity,
by viewing it as a result of the accumulation of the resonances
of clusters or sets of clusters of any size and shape.
The analysis of the complex conductivity of the $RL-C$
model in terms of resonances has been tackled in refs. [1, 8--10].
The present work provides a quantitative analysis
of the dielectric resonance spectrum of any cluster drawn on a regular lattice.
The case of the square lattice is considered for definiteness.
The setup of this paper is as follows.
Section 2 contains general results on
the resonance frequencies of arbitrary finite clusters and sets
of clusters, drawn on the square lattice,
as well as the associated resonance cross-sections.
Several applications are then presented in section 3,
including two coupled bonds, linear clusters,
arbitrary clusters of a given size (lattice animals),
and examples of self-similar fractal clusters.
Section 4 contains a short discussion.

\medskip
\noindent {\bf 2. GENERAL RESULTS}
\smallskip
\noindent {\bf 2.1. Definitions}
\smallskip
The aim of this section is to present a general approach
to the resonant dielectric response of clusters embedded in a regular lattice.
In the following a {\it cluster}
is any finite connected set of bonds drawn on a lattice,
and a {\it set of clusters} is a finite collection of such clusters.
We denote by $n_B$ the total number of bonds (links)
contained in a cluster or in a set of clusters,
and $n_S$ its total number of sites (vertices, or nodes).
For the sake of simplicity,
we restrict ourselves to the two-dimensional square lattice,
spanned by the unit vectors $\{\e_1$, $\e_2\}$.
We take the lattice spacing, i.e., the length of the bonds, as our length unit.

In order to study the dielectric response of a set of clusters,
we consider the geometry, shown in Figure 1,
of a rectangular sample of size $M\times N$,
between two straight parallel electrodes.
The bonds which belong to the set of clusters,
shown as thick lines and referred to as impurity bonds,
are assumed to have a complex, frequency-dependent conductance
(or admittance) $\sigma_1$, while the other bonds of the lattice,
shown as thin lines and referred to as matrix bonds,
have a different conductance $\sigma_0$.
The complex ratio $h$ of both conductances, defined in eq. (1.2),
will play a central r\^ole in the following.

Our starting point is the Kirchhoff equation for the electric potentials,
$$
\sum_{\y(\x)}\sigma_{\x,\y}(V_\x-V_\y)=\I_\x
.\eqno(2.1)
$$
We employ the following notations:
$\x=x_1\e_1+x_2\e_2=(x_1,x_2)$ is any site of the lattice,
and $\y(\x)$ denote the four nearest neighbouring sites of $\x$,
i.e., $\y=\x\pm\e_\mu$, with $\mu=1$ or 2.
The electric potential at site $\x$ is denoted by $V_\x$,
and $\sigma_{\x,\y}=\sigma_{\y,\x}$ is the conductance of the bond
joining the neighbouring sites $\x$ and $\y$.
Finally the source term $\I_\x$ is the current flowing from the generator
into the network through node $\x$;
it is non-vanishing only when $\x$ belongs to either electrode.

Both $V_\x$ and $\I_\x$ have to be solved from eq. (2.1),
with the boundary conditions $V_\x=0$ on the left electrode $(\x\in E_0)$,
and $V_\x=\V=M\E$ on the right one $(\x\in E_1)$,
with $\E$ being the uniform electric field applied to the sample.
The complex conductance (or admittance) between the electrodes reads
$$
Y=\frc{\I}{\V}
,\eqno(2.2)
$$
where $\I$ is the total current across the sample,
$$
\I=-\sum_{\x\in E_0}\I_\x=\sum_{\x\in E_1}\I_\x
.\eqno(2.3)
$$

\medskip
\noindent {\bf 2.2. The $RL-C$ model}
\smallskip
One situation of special interest is the $RL-C$ model
of inductive (and weakly dissipative) clusters in a dielectric matrix,
already mentioned in section 1.
The bonds of the clusters consist of an inductance $L$
in series with a weak resistance $R$,
while those of the matrix are perfect capacitances $C$.
The complex conductances at frequency $f=\omega/(2\pi)$ thus read
$$
\sigma_0=iC\omega,\qquad\sigma_1=\frc{1}{R+iL\omega}
.\eqno(2.4)
$$

Along the lines of refs. [1, 2, 9, 10],
we introduce the microscopic resonance frequency (plasmon frequency)
$$
\omega_0=\frc{1}{\sqrt{LC}}
,\eqno(2.5)
$$
and the quality factor
$$
Q=\frc{1}{R}\sqrt{\frc{L}{C}}=\frc{L\omega_0}{R}=\frc{1}{RC\omega_0}
,\eqno(2.6)
$$
which is a dimensionless measure of the dissipation rate.
We also introduce the reduced frequency
$$
y=\frc{\omega}{\omega_0}
,\eqno(2.7)
$$
so that
$$
h=\frc{1}{-LC\omega^2+iRC\omega}=\frc{1}{-y^2+iy/Q}
.\eqno(2.8)
$$

In the following, we shall mostly consider
the regime of a weak dissipation, corresponding to a large quality factor.
In this regime, we have therefore
$$
h\approx -\frc{1}{y^2}-\frc{i}{Qy^3}\qquad(Q\gg 1)
.\eqno(2.9)
$$

We have recalled in section 1 that $Y(h)$ is a rational function,
with alternating poles and zeros along the negative real axis.
Since the variable $h$ as given by eq. (2.9)
is very close to this negative real axis for $Q\gg 1$,
we can anticipate that the poles and zeros lying there will strongly affect
the frequency-dependent response of $RL-C$ clusters,
in the form of narrow resonances.
This is what we shall now demonstrate explicitly.

\medskip
\noindent {\bf 2.3. Resonance frequencies}
\smallskip
We define the resonances of a cluster, or a set of clusters,
as the values of $h$ such that the conductance $Y(h)$ vanishes,
in the limit of an infinitely large network.
This definition will be justified in section 2.4.

At a resonance, eq. (2.1) must have a non-trivial solution $V_\x$,
localised around the clusters, in the absence of sources.
This equation with $\I_\x=0$ can be recast as
$$
-(\Delta V)_\x=(1-h)\sum_{\y\in C(\x)} (V_\x-V_\y)
.\eqno(2.10)
$$
The notation $\y\in C(\x)$ means that the bond $(\x,\y)$
belongs to the set of clusters,
and $\Delta$ denotes the finite-difference Laplace operator
on the square lattice, defined as
$$
-(\Delta V)_\x=\sum_{\y(\x)} (V_\x-V_\y)
.\eqno(2.11)
$$

The solution of eq. (2.10) with appropriate decay properties at infinity reads
$$
\lambda V_\x=\sum_{\y\in C}\sum_{\z\in C(\y)}G_{\x,\y}(V_\y-V_\z)
,\eqno(2.12)
$$
where we have set
$$
\lambda=\frc{1}{1-h}
,\eqno(2.13)
$$
and where $G_{\x,\y}=G(\x-\y)$ is the Green's function of the Laplace operator
on the infinite square lattice.
Its main properties are recalled in the Appendix.

By expressing the consistency of eq. (2.12) for $\x$ being a site of the
cluster
set,
we arrive to the following characterisation of the resonances:
$\lambda$ has to be an eigenvalue of the square matrix $\M$,
of size $n_S\times n_S$, defined by
$$
\M_{\x,\y}=\sum_{\z\in C(\y)}(G_{\x,\y}-G_{\x,\z})
.\eqno(2.14)
$$
This matrix is not symmetric.
It can nevertheless be recast, using a bond representation,
in the form of a real symmetric matrix of size $n_B\times n_B$,
whose spectrum is manifestly real [10].

The $n_S$ eigenvalues of the matrix $\M$ lie in the range
$0\le\lambda\le 1$.
Only those different from the endpoints $(\lambda\ne 0$ and 1)
correspond to physical resonances.
We denote by $n_R$ the number of these non-trivial eigenvalues,
which we assume to be ordered as
$$
0 <\lambda_1\le\lambda_2\le\cdots\le\lambda_{n_R} < 1
.\eqno(2.15)
$$
Even for a single cluster,
there is no simple relationship between the numbers of sites,
of bonds, and of resonances, apart from the inequalities $n_R\le n_L<n_S$.
Furthermore, some of the eigenvalues $\lambda_a$ may be degenerate,
i.e., occur with a non-trivial multiplicity.

The spectrum of dielectric resonances $\lambda_a$ $(a=1,\cdots,n_R)$
of a given set of clusters thus only depends on the shape of the clusters,
and of their relative positions and orientations.
In the weak-dissipation regime of the $RL-C$ model, introduced in section 2.2,
the resonances occur at well-defined resonance frequencies
$f_a=\omega_a/(2\pi)$, given by
$$
\lambda_a=\frc{1}{1-h_a}=\frc{y_a^2}{y_a^2+1}
=\frc{\omega_a^2}{\omega_a^2+\omega_0^2},
\quad\hbox{i.e.,}\quad\omega_a=\omega_0\sqrt{\frc{\lambda_a}{1-\lambda_a}}
\qquad(Q\gg 1)
.\eqno(2.16)
$$

\medskip
\noindent {\bf 2.4. Resonance cross-sections}
\smallskip
We now turn to the determination of the analytic form
of the conductance $Y(h)$, in the regime of a weak dissipation $(Q\gg 1)$.
We restrict the analysis to the case where the linear sizes $M$ and $N$
of the sample are much larger than the diameter
of the set of clusters under consideration,
so that the resonances are very close to those determined in section 2.3,
corresponding to clusters embedded in an infinite lattice.
Section 4 contains a qualitative discussion of finite-size corrections.

Let $\M$ be the matrix associated with the clusters,
and let $\lambda_a$ $(a=1,\cdots,n_R)$ be its non-trivial eigenvalues.
We assume, for the sake of simplicity,
that all eigenvalues are non-degenerate.
For each eigenvalue $\lambda_a$, we denote by $h_a$, $y_a$, and $\omega_a$
the corresponding values of the various variables defined above,
the last two pertaining to the $RL-C$ model [see eq. (2.16)].
We also introduce the associated left and right eigenvectors
$L_a$ and $R_a$, with components $L_{a,\x}$ and $R_{a,\x}$
in the site representation (2.14).
These eigenvectors are supposed to be normalised in such a way that
$$
L_a.R_b=\sum_{\x\in C}L_{a,\x} R_{a,\x}=\delta_{a,b}\qquad(a,b=1,\cdots,n_R)
,\eqno(2.17)
$$
with $\delta_{a,b}$ being the Kronecker symbol.

We look for a solution of the Kirchhoff equation (2.1) in the form
$$
V_\x=\E x_1+W_\x
,\eqno(2.18)
$$
where the first term is the potential in the absence of the set of clusters,
with $x_1$ being the co-ordinate of the node $\x$ along the applied field $\E$,
and $W_\x$ is a perturbation of the potential,
localised around the clusters, and vanishing on the electrodes.
The Kirchhoff equations for the potential $W_\x$ on the clusters
can be recast as
$$
\lambda W_\x-\sum_{\y\in C}\M_{\x,\y}W_\y=\E\sum_{\y\in C}\M_{\x,\y} y_1
.\eqno(2.19)
$$
Whenever $\lambda$ comes close to one of the eigenvalues $\lambda_a$
of the matrix $\M$, the solution of eq. (2.19) diverges as
$$
W_\x\approx A(\lambda) R_{a,\x}
,\eqno(2.20)
$$
with $R_a$ being the corresponding right eigenvector.
The divergent prefactor $A(\lambda)$ is then determined
by multiplying the corresponding left eigenvector $L_a$ through eq. (2.19).
We thus obtain
$$
A(\lambda)\approx\frc{\lambda_a}{\lambda-\lambda_a}\E\L_a
\qquad(\lambda\to\lambda_a)
,\eqno(2.21)
$$
with
$$
\L_a=\sum_{\x\in C}x_1L_{a,\x}
.\eqno(2.22)
$$
This quantity is independent of the position of the set of clusters
inside the sample, owing to the identity $\sum_{\x\in C}L_{a,\x}=0$.
The total current across the sample is then evaluated as
$$
\I=\sigma_0
\left(N\E+\frc{h_a-1}{M}\sum_{(\x,\y)\in C}(x_1-y_1)(W_\x-W_\y)\right)
.\eqno(2.23)
$$

The conductance of the sample in the regime of a weak dissipation
can now be estimated by superposing the independent contributions
of all resonances to the intensity.
We are thus left with the following formula
$$
Y(h)\approx\frc{N}{M}\sigma_0
\left(1+\frc{1}{MN}\sum_{a=1}^{n_R}\gamma_a\,\frc{h_a}{h-h_a}\right)
,\eqno(2.24)
$$
with
$$
\gamma_a=\frc{\L_a\R_a}{\lambda_a(1-\lambda_a)}
,\eqno(2.25)
$$
and
$$
\R_a=\sum_{(\x,\y)\in C}(x_1-y_1)(R_{a,\x}-R_{a,\y})
.\eqno(2.26)
$$

Eq. (2.24) is our main result.
The prefactor $(N/M)\sigma_0$ is the conductance of the matrix sample,
in the absence of the clusters.
To this background response
are superposed the $n_R$ resonances of the set of clusters.
In agreement with the general properties recalled in section 1,
each resonance shows up as a doublet,
consisting of a pole situated at $h=h_a$
and a zero situated at $h\approx h_a\bigl(1-\gamma_a/(MN)\bigr)$.

The strength of each resonance is measured by the distance
between the pole and the zero, i.e., by the residue $\gamma_ah_a/(MN)$.
It is inversely proportional to the area $MN$ of the sample,
and proportional to the parameter $\gamma_a$,
that we call the {\it cross-section} of the resonance.
The latter is indeed interpreted as the area
of the region around the clusters where the perturbation
$W_\x$ of the potential at resonance takes appreciable values.

The general result (2.24) can be made more explicit in the $Q\gg 1$ regime
of the $RL-C$ model, introduced in section 2.2.
In this situation, it is advantageous to consider the impedance $Z=1/Y$
of the sample.
The latter quantity reads, in terms of the reduced frequency $y$ of eq. (2.7),
$$
Z\approx -i\frc{M}{NC\omega_0y}+\frc{1}{2N^2C\omega_0}\sum_{a=1}^{n_R}
\gamma_a\,\frc{1/(2Q)-i(y-y_a)}{(y-y_a)^2+1/(4Q^2)}
.\eqno(2.27)
$$

The real (dissipative) part of the impedance exhibits narrow resonance peaks,
having Lorentzian line shapes, with a common absolute width
$$
\Delta\omega=\frc{\omega_0}{2Q}
.\eqno(2.28)
$$
Both the maximal value at resonance
$$
(\re Z)_\max\approx\frc{\gamma_a Q}{N^2C\omega_0}
\eqno(2.29)
$$
and the area under the resonance peak
$$
{\cal A}_a=\int_{\omega\approx\omega_a}\re Z\,\d\omega\approx\frc{\pi\gamma_a
}{2N^2C}
\eqno(2.30)
$$
depend on the clusters and on the resonance under consideration
only through the cross-section $\gamma_a$.

\medskip
\noindent {\bf 2.5. Duality symmetry}
\smallskip
Duality is one of the key concepts of the theory of planar graphs
(see ref. [11] for an overview).
It has far-reaching consequences in two-dimensional statistical physics
(see ref. [12] for a review).
The applications of duality to random resistor networks
have been reviewed in ref. [1].
The dual of the network of Figure 1 has its electrodes along
the horizontal sides ($M$ and $N$ are interchanged),
and the dual clusters are obtained by drawing a dual bond
across each bond of the original ones.
Figure 2 shows some illustrative examples of clusters with their duals.
Case (c) demonstrates that the dual of a connected cluster
may be disconnected.

The duality symmetry of the square lattice implies the identity
$$
Y(h)\;\w Y(\w h)=1
\eqno(2.31)
$$
between the conductances of the original network and of its dual,
with bond conductances related through
$\w\sigma_0=1/\sigma_0$, $\w\sigma_1=1/\sigma_1$, so that $\w h=1/h$.
By inserting the result (2.24) into the identity (2.31),
we obtain the following predictions.
The resonances of the dual set of clusters are located at the dual positions,
$$
\quad\w h_a=1/h_a,\quad\hbox{i.e.,}\quad\w\lambda_a=1-\lambda_a,
\quad\hbox{or}\quad\w\omega_a=\omega_0^2/\omega_a
,\eqno(2.32)
$$
and any pair of dual resonances have identical cross-sections,
$$
\w\gamma_a=\gamma_a
.\eqno(2.33)
$$

As an illustration we close up this general section by giving in Table 1
the positions $\lambda_a$ of the resonances,
and the associated cross-sections $\gamma_a$,
for the three clusters shown in Figure 2.
Cluster (a) is a generic example.
Cluster (b) is self-dual, i.e., isometric to its dual,
so that its resonances come in dual pairs.
Cluster (c) has two peculiarities, namely it possesses a closed loop
and its dual is not connected.

\medskip
\noindent {\bf 3. APPLICATIONS}
\smallskip
\noindent {\bf 3.1. One bond}
\smallskip
Consider first the simplest of all clusters, consisting of only one bond,
joining the origin $\x=\0$ to $\x'=\e_\mu$.
The bond is parallel to the applied field for $\mu=1$,
and perpendicular to it for $\mu=2$.
The corresponding matrix $\M$ only involves two values of the Green's function,
$G_0$ and $G_1$, given by eq. (A.5).
We thus get
$$
\M=\frc{1}{4}\pmatrix{1&-1\cr-1&1}
,\eqno(3.1)
$$
with eigenvalues
$$
\lambda_1=1/2,\qquad\lambda_2=0
.\eqno(3.2)
$$
Only $\lambda_1$ leads to a resonance, lying at
$$
h_1=-1
,\eqno(3.3)
$$
i.e., $\omega=\omega_0$ for the $RL-C$ model.
The associated cross-section reads
$$
\gamma_1=4\delta_{\mu,1}
,\eqno(3.4)
$$
with $\delta_{\mu,1}$ being the Kronecker symbol.
We thus have $\gamma_1=4$ for $\mu=1$
(the bond is parallel to the applied field),
and $\gamma_1=0$ for $\mu=2$ (the bond lies along an equipotential line).

The latter property is quite general.
Clusters consisting only of vertical bonds,
perpendicular to the applied field, do not perturb it at all.
Their resonances thus have vanishing cross-sections.

\medskip
\noindent {\bf 3.2. Two bonds}
\smallskip
Consider now two bonds, in arbitrary relative position and orientation,
namely one joining the points $\x=\0$ and $\x'=\e_\mu$,
and one joining the points $\y=(y_1,y_2)$ and $\y'=\y+\e_\nu$,
with $\mu,\nu=1$ or 2.

The corresponding matrix $\M$ is
$$
\M=\pmatrix{
1/4 & -1/4 & G_{\x,\y}-G_{\x,\y'} & G_{\x,\y'}-G_{\x,\y}\cr
-1/4 & 1/4 & G_{\x',\y}-G_{\x',\y'} & G_{\x',\y'}-G_{\x',\y}\cr
G_{\x,\y}-G_{\x',\y} & G_{\x',\y}-G_{\x,\y} & 1/4 & -1/4\cr
G_{\x,\y'}-G_{\x',\y'} & G_{\x',\y'}-G_{\x,\y'} & -1/4 & 1/4\cr}
.\eqno(3.5)
$$
Its eigenvalues read
$$
\lambda_1=1/2+g,\qquad\lambda_2=1/2-g,\qquad\lambda_3=\lambda_4=0
,\eqno(3.6)
$$
with
$$
\eqalign{
g&=G_{\x,\y}-G_{\x,\y'}-G_{\x',\y}+G_{\x',\y'}\cr
&=G_{\mu,\nu}(\y)\equiv G(\y)-G(\y+\e_\nu)-G(\y-\e_\mu)+G(\y+\e_\nu-\e_\mu).
}
\eqno(3.7)
$$
Only $\lambda_1$ and $\lambda_2$ lead to resonances, lying at
$$
h_1=-\frc{1-2g}{1+2g},\qquad h_2=-\frc{1+2g}{1-2g}=\frc{1}{h_1}
.\eqno(3.8)
$$
The associated cross-sections are
$$
\gamma_1=\frc{2(\delta_{\mu,1}+\delta_{\nu,1})^2}{1-4g^2},\qquad
\gamma_2=\frc{2(\delta_{\mu,1}-\delta_{\nu,1})^2}{1-4g^2}
.\eqno(3.9)
$$

In the limit where both bonds are infinitely far apart $(g=0)$,
we obtain two degenerate resonances at $h=-1$.
If the bonds are at a large but finite distance
$(\vert\y\vert\gg 1)$, the estimate (A.6) shows that the coupling
between both resonances is approximately
$$
g\approx\frc{1}{2\pi}\frc{\partial^2}{\partial y_\mu\partial
y_\nu}\ln\vert\y\vert
=\frc{1}{2\pi\vert\y\vert^2}\left(\delta_{\mu,\nu}-2\frc{y_\mu
y_\nu}{\vert\y\vert^2}\right)\ll 1
.\eqno(3.10)
$$
The resonances are thus symmetrically shifted by a small amount,
$$
h_1,h_2\approx -1\pm 4g
,\eqno(3.11)
$$
hence, in the case of the $RL-C$ model,
$$
\omega_1,\omega_2\approx\omega_0(1\mp 2g)
.\eqno(3.12)
$$

\medskip
\noindent {\bf 3.3. Linear clusters}
\smallskip
Another case of interest is that of linear clusters.
Consider a horizontal linear cluster consisting of $n_B=n$ consecutive bonds,
joining the sites $x\e_1$, with $x=0,\cdots,n$.
The entries of the corresponding matrix $\M$
only involve the Green's function $G_x$, with the notation of the Appendix.
They are symmetric for generic values of the site labels,
$$
\M_{x,y}=2G_{x-y}-G_{x-y-1}-G_{x-y+1}\equiv F(x-y)
\qquad(y=1,\cdots,n-1)
,\eqno(3.13)
$$
together with the non-symmetric boundary values
$$
\M_{x,0}=G_x-G_{x-1},\qquad\M_{x,n}=G_{x-n}-G_{x+1-n}
.\eqno(3.14)
$$

The positions of the resonances and the corresponding cross-sections
can be determined analytically for the first few values of the size $n$:

\noindent $\bullet$ $n=1$.
This is the one-bond case, investigated in section 3.1.

\noindent $\bullet$ $n=2$.
This is a special case of the two-bond case investigated in section 3.2,
corresponding to $\mu=\nu=1$ and $\y=\e_1$, hence
$$
g=2G_1-G_2=1/2-2/\pi
.\eqno(3.15)
$$
The general results (3.6, 8, 9) yield
$$
\matrix{
\lambda_1=1-2/\pi=0.36338,\hfill&\lambda_2=2/\pi=0.63662,\hfill&\lambda_3=0,\cr
h_1=-2/(\pi-2)=-1.7519,\qquad&h_2=1-\pi/2=-0.57080,&\,\cr
\gamma_1=\pi^2/(\pi-2)=8.6455,\hfill&\gamma_2=0.\hfill&\,\cr
}
\eqno(3.16)
$$

\noindent $\bullet$ $n=3$.
The matrix
$$
\M=\pmatrix{-G_1&2G_1-G_2&2G_2-G_1-G_3&G_3-G_2\cr G_1&-2G_1&2G_1-G_2&G_2-G_1\cr
G_2-G_1&2G_1-G_2&-2G_1&G_1\cr G_3-G_2&2G_2-G_1-G_3&2G_1-G_2&-G_1}
\eqno(3.17)
$$
has eigenvalues
$$
\eqalign{
\lambda_1&=(7-16/\pi-w)/4=0.28216,\qquad\lambda_2=8/\pi-2=0.54648,\cr
\lambda_3&=(7-16/\pi+w)/4=0.67136,\qquad\lambda_4=0,
}\eqno(3.18)
$$
with $w=\sqrt{384/\pi^2-224/\pi+33}=0.77841$, hence
$$
\eqalign{
h_1&=-(16/\pi+w-3)/(7-16/\pi-w)=-2.54411,\cr
h_2&=-(3-8/\pi)/(8/\pi-1)=-0.82990,\cr
h_3&=-(16/\pi-w-3)/(7-16/\pi+w)=-0.48951.
}\eqno(3.19)
$$
We have also evaluated the associated cross-sections analytically, namely
$$
\eqalign{
\gamma_1&
=\frc{64(1+w)(1-3/\pi)}{w(16/\pi+w-7)(24/\pi-w-7)(16/\pi+w-3)}=14.633,\cr
\gamma_2&=0,\cr
\gamma_3&
=\frc{64(1-w)(1-3/\pi)}{w(-16/\pi+w+7)(24/\pi+w-7)(16/\pi-w-3)}=0.16406.
}\eqno(3.20)
$$

\noindent $\bullet$ $n\gg 1$.
The resonance spectrum of very long linear clusters
can also be investigated as follows.
The bulk matrix entries (3.13) only involve an even function $F(x-y)$
of the distance between nodes along the cluster.
As a consequence, the matrix $\M$ is approximately a Toeplitz matrix,
which is diagonalisable by means of the Fourier transformation.
The eigenvalue $\lambda$ is given in terms of a wavevector (or momentum) $q$,
by a dispersion relation of the form
$$
\lambda(q)=\sum_{x=-\infty}^{+\infty}F(x)e^{-iqx}
.\eqno(3.21)
$$
The expression (A.2) of the Green's function leads to
$$
\lambda(q)=\int_0^{2\pi}\frc{\d p}{2\pi}\frc{1-\cos q}{2-\cos q-\cos p}
=\sqrt\frc{1-\cos q}{3-\cos q}={\vert\sin(q/2)\vert\over\sqrt{1+\sin^2(q/2)}}
.\eqno(3.22)
$$

For a large but finite linear cluster, the wavevector $q$ is quantised.
It assumes $n$ discrete values in the range $0<q<\pi$, approximately given by
$$
nq\approx a\pi\qquad(a=1,\cdots,n)
.\eqno(3.23)
$$

The positions of the resonances are therefore asymptotically distributed
according to the smooth density
$$
\rho(\lambda)=\frc{1}{\pi}\;\frc{\d q}{\d\lambda}
=\frc{2}{\pi(1-\lambda^2)\sqrt{1-2\lambda^2}}
.\eqno(3.24)
$$
Two limiting situations deserve our attention

\noindent $\star$ In the long-wavelength regime $(q\to 0)$,
which corresponds to low frequencies in the $RL-C$ model,
we have the linear dispersion law $\lambda(q)\approx q/2$,
hence the constant density of resonances
$$
\rho(\lambda)=\frc{2}{\pi}\qquad(\lambda\to 0)
.\eqno(3.25)
$$
The scaling behaviour of the resonance spectrum in this regime
will be investigated from a more general viewpoint in section 3.6.

\noindent $\star$ In the opposite limit $(q\to\pi)$,
$\lambda$ approaches a non-trivial maximal value
$$
\lambda_L=1/\sqrt{2}=0.70711
.\eqno(3.26)
$$
Near this upper band-edge of the dispersion law,
the density of resonances (3.24) exhibits an
inverse-square-root van-Hove divergency.

The resonances of linear clusters thus extend over
the range $0<\lambda<\lambda_L$, i.e., $-\infty<h<h_L$,
with $h_L=-(\sqrt 2-1)$.
The resonance spectra of the first 14 linear clusters are shown in Figure 3.

We close up with a few words about the resonance
cross-sections of linear clusters.
First, the vanishing of $\gamma_2$ for the linear clusters
with $n=2$ and $n=3$ [see respectively eqs. (3.16) and (3.20)]
is in fact quite general.
Indeed, because of the left-right symmetry of linear clusters,
the eigenvectors of the matrix $\M$ have a definite parity.
We have $R_{a,n-k}=(-1)^a R_{a,k}$,
and a similar symmetry property for the $L_a$, hence $\R_a=0$ for $a$ even.
The cross-section $\gamma_a$ therefore vanishes for every even $a$,
and every $n$.
Furthermore, in the long-wavelength regime
$(q\ll 1$, i.e., $\lambda\ll 1$ or $a\ll n)$,
the eigenvectors $L_{a,k}$ and $R_{a,k}$ are
given by superpositions of plane waves $\exp(\pm iqk)$,
and thus approximately uniformly extended over the cluster.
Hence both amplitudes $\L_a$ and $\R_a$ scale as the size $n$,
so that we obtain the following scaling form of the cross-sections
$$
\gamma_a\approx n^2 A_a\qquad(a\ \,\hbox{odd},\;n\gg 1)
,\eqno(3.27)
$$
where the $A_a$ are numbers of order unity.
This scaling law will be given a simple interpretation in section 4.

\medskip
\noindent {\bf 3.4. Lattice animals}
\smallskip
A lattice animal (or an animal, for short)
is by definition any connected cluster drawn on a given regular lattice.
The problem of lattice animals has been an active field
of statistical mechanics,
one of the main motivations arising from cluster-expansion techniques
for percolation and other problems in lattice statistics [13].
Just as percolation, the lattice-animal problem can be formulated
either in terms of site occupation, or of bond occupation.
The problem of bond animals is more suited for conduction properties.
It consists in considering, with equal statistical weights,
all clusters drawn on the lattice,
made of a given number $n_B=n$ of bonds.
The questions which have attracted most attention so far concern
the asymptotic $(n\to\infty)$ behaviour of the total number of animals,
and of geometrical characteristics, such as their mean radius of gyration.

We have investigated the resonance spectra of animals
drawn on the square lattice.
To do so, we have generated all lattice animals, up to $n_\max=11$,
by adapting to the bond problem [10]
an algorithm known for the site problem [14],
and evaluated numerically the resonances of each animal.
The total numbers $N_n$ of animals of size $n$ are listed in Table 2.
This quantity grows asymptotically as
$$
N_n\sim\frc{\mu^n}{n}
,\eqno(3.28)
$$
where the universal $1/n$ power-law is an exact result in two dimensions [15],
while the connectivity constant, here $\mu\approx 5.208$,
depends on the underlying lattice.

Figure 4 shows histogram plots of the resonance spectra of all animals
consisting of 7, 9, and 11 bonds.
Several peaks are clearly visible.
Two of the most salient ones are shown on the plots,
namely the upper band-edge $\lambda_T$ (3.26) of the linear clusters,
and the upper band-edge of the $T$-shaped clusters
(i.e., the long linear clusters decorated by one bond
branching perpendicularly at any place),
that we have evaluated numerically to be at $\lambda_T=0.83781$.
\vfill\eject
\noindent {\bf 3.5. Connections with the random binary model}
\smallskip
We now come to the question addressed in section 1,
on the relationship between the singularities of the conductivity
of the random binary network in the complex $h$-plane,
and the resonances of isolated clusters.
We restrict again the analysis to the case of the square lattice.

A systematic way of addressing this question
is to consider the regime of very diluted impurities,
$$
c=1-p\ll 1
.\eqno(3.29)
$$
The conductivity has been shown [16--18] to
admit a power-series expansion in $c$, of the form
$$
\Sigma(h,c)=\sigma_0\Bigl(1+b_1(h)c+b_2(h)c^2+\cdots\Bigr)
,\eqno(3.30)
$$
where the coefficients $b_k(h)$ only depend on
the lattice under consideration and on the conductance ratio $h$.
It is worth noticing that the series expansion (3.30)
is different from the weak-disorder expansion
of the conductivity as a power series in the successive cumulants
of an arbitrary (smooth) distribution of the bond conductances,
investigated in ref. [19].

The coefficients of the power-series expansion (3.30)
are constrained by the duality symmetry of the square lattice.
Indeed, by inserting this expansion into the thermodynamic
version of the identity (2.31) [1],
$$
\Sigma(h,c)\Sigma(1/h,c)=\sigma_0^2
,\eqno(3.31)
$$
independently of the concentration $c$ of impurity bonds,
we can derive an infinity of relations for the functions $b_k(h)$, namely
$$
b_1(h)+b_1(1/h)=0,\qquad b_2(h)+b_2(1/h)=b_1(h)^2
,\eqno(3.32)
$$
and so on.

\noindent $\bullet$
The first coefficient $b_1(h)$ only involves one-impurity effects.
It is therefore correctly predicted by the EMA formula (1.6), which yields
$$
b_1(h)=b_1^\EMA(h)=\frc{2(h-1)}{h+1}
.\eqno(3.33)
$$

\noindent $\bullet$
The higher-order coefficients of the expansion (3.30) are far more difficult
to calculate.
The determination of $b_2(h)$, which contains two-impurity effects,
has been performed on the cubic lattice [16], on the square lattice [17],
and on the hypercubic lattice in any dimension $D$,
by means of a two-impurity improved EMA-like scheme [18].
Coming back to the square lattice,
the most appropriate expression of $b_2(h)$ for the present purpose reads [18]
$$
b_2(h)=\frc{1}{2}b_1(h)^2
+b_1(h)^3\sum_{(\x,\mu)\ne(\0,1)}
\frc{\bigl[G_{1,\mu}(\x)\bigr]^2}{1-\bigl[b_1(h)G_{1,\mu}(\x)\bigr]^2}
-b_1(h)^4\sum_{\x\ne\0}
\frc{\bigl[G_{1,1}(\x)\bigr]^3}{1-\bigl[b_1(h)G_{1,1}(\x)\bigr]^2}.
\eqno(3.34)
$$
This result is obviously more intricate than the EMA prediction
$$
b_2^\EMA(h)=\frc{4h(h-1)^2}{(h+1)^3}=\frc{b_1(h)^2\bigl(b_1(h)+2\bigr)}{4}
.\eqno(3.35)
$$
For instance, in the case of insulating impurity bonds $(h=0)$,
eq. (3.34) yields $b_2(0)=-0.210749$, while $b_2^\EMA(0)$ vanishes.

First of all, the results (3.33, 34) obey the identities (3.32).
This is obvious for $b_1$, while more involved symmetry arguments
are needed to check this property for $b_2$, along the lines of ref. [19].

The result (3.33) shows that the coefficient $b_1(h)$ has a pole
at $h=-1$, corresponding to the resonance (3.3) of the one-bond problem.
Similarly, eq. (3.34) shows that $b_2(h)$ has an infinity of poles,
situated at $b_1(h)=\pm 1/G_{\mu,\nu}(\x)$,
corresponding to all resonances of the two-bond configurations,
determined in section 3.2.
Indeed eq. (3.8) can be recast as $b_1(h)=\pm 1/g$.

This observation can be generalised to the following rule [10, 18].
For any $k\ge 2$,
the coefficient $b_k(h)$ of the power-series expansion (3.30) has an infinity
of poles along the negative real $h$-axis,
corresponding to all resonances of all sets of clusters
consisting altogether of $k$ bonds.
This shows in particular that the conductivity of the random binary model
is singular along the whole negative real axis of the complex $h$-variable.
Indeed our investigations of the resonance spectra of linear clusters
and of lattice animals demonstrate
a dense accumulation on the negative real axis.

\medskip
\noindent {\bf 3.6. Fractal clusters}
\smallskip
This last section is devoted to the resonances of large clusters
drawn on the square lattice, including fractal ones.
It turns out that the spectra of large fractal clusters
generically exhibit a scaling power-law behaviour
in the $\lambda\ll 1$ regime, i.e., for $a\ll n_R$.
We shall explain this phenomenon in a heuristic way,
making use of our intuition of the low-frequency response
of $RL-C$ clusters, where $\lambda\approx(\omega/\omega_0)^2$.

Consider a large but finite patch of a fractal cluster,
of diameter $\ell$, embedded in the square lattice.
The numbers of its sites, bonds, and resonances scale as
$$
n_S\sim n_B\sim n_R\sim\ell^{D_F}
,\eqno(3.36)
$$
where $D_F\le 2$ is the Hausdorff (or fractal) dimension of the cluster.

The investigation of the resonances of linear clusters,
performed in section 3.3, has shown that the resonances for $\lambda\ll 1$
are characterised by coherent long-wavelength excitations along the cluster.
We propose to generalise this picture to clusters of arbitrary shape,
and to assert that the lowest resonance is generically given by
$$
\lambda_\min\approx
(\omega_\min/\omega_0)^2\sim\frc{1}{L(\ell)C(\ell)\omega_0^2}
,\eqno(3.37)
$$
where $L(\ell)$ and $C(\ell)$ are the effective inductance
and the effective capacitance of the cluster, considered as a whole:

\noindent $\bullet$
The effective inductance $L(\ell)$ of a cluster obeys
the same scaling behaviour as its end-to-end resistance,
$$
L(\ell)\sim R(\ell)\sim\ell^{t/\nu}
,\eqno(3.38)
$$
where $t/\nu$ is the usual notation coming from
finite-size scaling theory for the percolation problem (see e.g. ref. [1]).

\noindent $\bullet$
To the contrary, the capacitance $C(\ell)$ of the cluster is
assumed to be insensitive to its internal structure.
In two dimensions it is therefore independent of $\ell$, so that we obtain
$$
\lambda_\min\sim\ell^{-t/\nu}
.\eqno(3.39)
$$

This scaling law can be extended to all resonances such that $\lambda\ll 1$,
i.e., $a\ll n_R$, by expressing that in this regime $\lambda_a$ only depends
on the label $a$ through the dimensionless combination $a/n_R\ll 1$.
We thus get
$$
\lambda_a\sim (a/n_R)^\zeta\quad(a\ll n_R),
\quad\hbox{with}\qquad\zeta=\frc{t}{\nu\,D_F}
.\eqno(3.40)
$$
An equivalent statement consists in writing the scaling law
$$
\rho(\lambda)\sim\lambda^{-1+1/\zeta}\qquad(\lambda\to 0)
\eqno(3.41)
$$
for the density of resonances of the infinite fractal structure.

For linear clusters, we have $t/\nu=D_F=1$,
so that $\zeta=1$, in agreement with the analytical results (3.23, 24).

We have also investigated three examples of deterministic
self-similar fractal clusters,
which we call the Worm, the Cross and the Gasket.
Their iterative construction rules are shown in Figure 5,
while the resulting clusters are shown in Figure 6.
At each step of the construction, the number of bonds
is increased by a constant factor $a$,
while the diameter of the cluster (in units of the bond length)
is multiplied by a scaling factor $b$.
After $k$ steps of iteration,
the fractal cluster of the $k$-th generation thus consists
of $n_B\sim a^k$ bonds, while its diameter scales as $\ell\sim b^k$.
As a consequence, the fractal dimension of these clusters is
$$
D_F=\ln a/\ln b
.\eqno(3.42)
$$
The exponent $t/\nu$ of the end-to-end resistance is trivial
in the first two examples, which possess no closed loops.
Its value for the Gasket is known for long,
and rederived e.g. in the review [1].
The values of the various scaling factors and exponents
for the three fractal clusters are listed in Table 3.
We have evaluated numerically the resonance spectra of successive generations
of these three fractal clusters, by diagonalising the associated $\M$ matrices,
up to a maximal scale $k_\max$ corresponding to a few thousand bonds.
Figure 7 shows logarithmic plots of the resonance positions
against $a/n_R\ll 1$, for the largest two scales of each type of cluster.
Power laws clearly show up, with exponents $\zeta$ in agreement
with the prediction (3.40), listed in Table 3.

\medskip
\noindent {\bf 4. DISCUSSION}
\smallskip
We have investigated the resonances
which show up in the conductivity of a connected cluster,
and of a finite set of clusters,
drawn on the square lattice and embedded in a large rectangular sample.
Our central result (2.24) shows that the conductivity
is characterised by a finite number $n_R$ of resonances,
located at well-defined negative real values $h_a$ of the ratio $h$
of the conductivities of both phases.
This formalism is chiefly aimed at describing
the resonant dielectric response of clusters within the $RL-C$ model.

The strength of each resonance is measured by a cross-section $\gamma_a$,
given by eq. (2.25).
The cross-section is naturally interpreted as the area of the sample
over which the cluster significantly perturbs the applied field $\E$
at resonance.
A slightly different and complementary viewpoint is as follows.
Each factor $\L_a$ or $R_a$ of eq. (2.25) has the dimension of a length.
It can be interpreted as a measure
of the strength of the dipole induced on the cluster by the applied field,
so that the cross-section $\gamma_a$ is proportional to the square of the
induced dipole.
The result (3.20) shows however
that there is no simple algebraic relation between $\L_a$ and $\R_a$
in general, and that $\gamma_a$ is not mathematically a perfect square.

This interpretation of the cross-section sheds some light on several
aspects of this work.
First, the coupling constant $g$ (3.10)
between the resonances of two distant bonds
coincides with the interaction energy of two dipoles $\D$ and $\D'$
in two-dimensional electrostatics, namely
$$
E=-(\D.\nabla)(\D'.\nabla')\frc{\ln r}{2\pi}
=\frc{1}{2\pi}\left(\frc{\D.\D'}{r^2}-2\frc{(\D.\r)(\D'.\r)}{r^4}\right)
.\eqno(4.1)
$$
The $n^2$-law (3.27) for the cross-sections of long linear clusters
is also simply understood as follows.
For $\lambda\ll 1$, i.e., at low frequency for the $RL-C$ model,
the clusters respond coherently,
so that the induced dipole is proportional to the size $n$.
Our interpretation of the cross-section also allows a qualitative discussion
of the finite-size corrections to our predictions.
The most important of these effects consists in a shift
of the spectrum of resonances.
For a large but finite sample of size $M\times N$,
the elements of the matrix $\M$
differ from those corresponding to the infinite square lattice
by terms of order $1/M^2$ and $1/N^2$,
proportional to the interaction energy (4.1) between the dipole induced
on the cluster and its mirror images with respect to the boundaries
of the sample.
The resonances are therefore shifted
by a small amount of order $1/M^2$ and $1/N^2$,
with respect to their theoretical positions $h_a$.

The present analysis can be extended to the problem
of a {\it coloured cluster}, consisting of different kinds of metallic bonds.
Assume that each bond of the cluster has a conductance
$\sigma_{\x,\y}=\sigma_0 h_{\x,\y}$,
where the $h_{\x,\y}$ are given arbitrary complex numbers.
A generalisation of eq. (2.10) shows that the resonance condition now reads
$$
\det\bigl(\1-\N)=0
,\eqno(4.2)
$$
where $\N$ is an $n_S\times n_S$ matrix, defined by
$$
\N_{\x,\y}
=\sum_{\z\in C(\y)}\bigl(1-h_{\y,\z}\bigr)\bigl(G_{\x,\y}-G_{\x,\z}\bigr)
.\eqno(4.3)
$$
The condition (4.2) yields a polynomial relation between the $h_{\x,\y}$.
A simple illustration is given by the linear cluster consisting
of two neighbouring bonds with conductance ratios $h_1$ and $h_2$.
We thus obtain the relation
$$
(1+h_1)(1+h_2)=(1-4/\pi)^2(1-h_1)(1-h_2)
.\eqno(4.4)
$$
If all bonds of the cluster are identical, i.e., $h_{\x,\y}=h$,
we have $\N=\M/\lambda$.
The eigenvalue criterion is thus recovered.
In our example, eq. (4.4) for $h_1=h_2$ yields back the result (3.16).

We have also investigated the small-$\lambda$ region of the resonance spectra
of large clusters, including fractal ones, embedded in the square lattice.
The scaling laws (3.40), (3.41), established in a heuristic way,
have been checked analytically by means of the dispersion law
pertaining to linear clusters,
and numerically on several examples of self-similar fractals.
We would like to emphasise that these scaling laws
cannot be generalised in a straightforward way
to clusters embedded in a higher-dimensional lattice.
This can be demonstrated explicitly in the case of linear clusters.
On the hypercubic lattice in dimension $D$,
the dispersion relation for a small wavevector $(q\ll 1)$ reads
$$
\lambda(q)\approx\int\!\frc{\d^{D-1}\p}{(2\pi)^{D-1}}\frc{q^2}{q^2+\p^2}
\sim\left\{\matrix{
q^{D-1}\hfill&(D<3),\cr
q^2\ln(\Lambda^2/q^2)\hfill&(D=3),\cr
q^2\hfill&(D>3).\cr
}\right.
\eqno(4.5)
$$
This result implies the existence of an upper critical dimension $D_c=3$,
at least in the case of linear clusters.

It should also be noticed that
the present results only concern the dielectric resonances of finite clusters
embedded in an infinite matrix, which consists of a regular lattice.
The resonance spectra of deterministic models
for fractal structures, such as the incipient percolation cluster,
investigated in refs. [20, 21], have very different characteristics.
In the latter case both phases have comparable sizes and geometries,
so that the notions of cluster and matrix are no more pertinent.
The spectra of such model systems are generically themselves fractal.
They are supported by a Cantor set of the negative real $h$-axis,
instead of the whole axis in the present situation.

To close up, we come back to our initial motivation,
namely a better understanding of the analytic structure
of the conductivity of the random binary model,
both in the conductance ratio $h$ and in the concentration variable $p$ or $c$.
First of all, and from a qualitative viewpoint,
the present analysis confirms that the cut of the conductivity
in the complex $h$-plane can be viewed as the accumulation of the resonances
of clusters and sets of clusters with any size and shape,
and that it extends over the whole negative real axis.
The more quantitative analysis of section 3.5
shows that every coefficient $b_k(h)$ of the expansion (3.30)
of the conductivity has itself a countable infinity of poles,
associated with the resonances of all embeddings, connected or not,
of $k$ bonds in the square lattice.
The present work leaves several open questions,
such as e.g. a more accurate form for the Lifshitz tail (1.9).
It however shows in a very suggestive way how intricate the exact conductivity
of the binary model can be on a finite-dimensional lattice,
especially as compared to the simple and very commonly used EMA formula.
\bigskip
\noindent {\bf Acknowledgements}
\smallskip
It is a pleasure for us to thank F. Brouers, B. Hesselbo, J.M. Laugier,
Th.M. Nieuwenhuizen, J.J. Niez, Y.P. Pellegrini, B. Souillard, B. T\'edenac,
and Z. Randriamanantany for fruitful discussions during our investigations.
J.-M. Normand is gratefully acknowledged for a careful reading
of the manuscript.

This work has been partly supported by DRET contract number 92490 of
12.02.1993.

\vfill\eject
\noindent {\bf APPENDIX}
\smallskip
In this Appendix we summarise the main properties of the Green's function
of the finite-difference Laplace operator $\Delta$ on the square lattice,
which are useful in the body of this paper.
The reader is referred to ref. [22] for a more comprehensive exposition.

The Green's function $G_{\x,\y}=G(\x-\y)$ is by definition a solution of
$$
-\Delta G(\x)=\delta_{\x,\0}
,\eqno({\rm A}.1)
$$
with $\delta_{\x,\0}$ being the Kronecker symbol.
The difference equation (A.1) has a unique solution
with all required symmetries, up to an additive constant,
which we fix by setting $G(\0)=0$.
We obtain by Fourier transformation
$$
G_{\x,\y}=G(\x-\y)=G(x_1-y_1,x_2-y_2)
=\int_B\frc{\d^2\p}{(2\pi)^2}\frc{e^{i\p.(\x-\y)}-1}{K(\p)}
,\eqno({\rm A}.2)
$$
where the double integral runs over the first Brillouin zone $B$
$(-\pi<p_1, p_2<\pi)$, and where
$$
K(\p)=2(2-\cos p_1-\cos p_2)
\eqno({\rm A}.3)
$$
is the Fourier transform of $(-\Delta)$.

The values of $G(\x)$ along the diagonals, namely for $m=\pm n$,
can be evaluated explicitly from the representation (A.2)
by means of elementary integrals.
One thus obtains
$$
G(\pm n,\pm n)
=-\frc{1}{\pi}\left(1+\frc{1}{3}+\cdots+\frc{1}{2n-1}\right)\qquad(n\ge 1)
.\eqno({\rm A}.4)
$$
The values of $G(\x)$ all over the square lattice
can then be determined recursively from eq. (A.1),
starting from the knowledge of the values (A.4) [22].

Introducing the short-hand notation $G_m=G(\pm m,0)=G(0,\pm m)$
for the values of the Green's function along the co-ordinate axes, we have
$$
G_0=0,\quad G_1=-1/4,\quad G_2=2/\pi-1=-0.36338,\quad G_3=12/\pi-17/4=-0.43028
,\eqno({\rm A}.5)
$$
and so on.

Finally, the Green's function admits the long-distance expansion
$$
G(\x)\approx-\frc{1}{2\pi}\left[\ln\vert\x\vert+\left(\frc{3}{2}\ln
2+\euler\right)+\frc{8x_1^2x_2^2-\vert\x\vert^4}
{12\vert\x\vert^6}+\cdots\right]\qquad(\vert\x\vert\gg 1)
.\eqno({\rm A}.6)
$$
The leading isotropic logarithmic term is nothing but the Green's function of
the differential Laplace operator in the plane,
which represents e.g. the potential of a point charge
in two-dimensional electrostatics.
The finite part involves Euler's constant $\euler=0.57721$.
The first anisotropic correction, due to lattice effects,
is of relative order $1/\vert\x\vert^2$.
\vfill\eject
{
\parindent 0pt
{\bf REFERENCES}
\bigskip

[1] J.P. Clerc, G. Giraud, J.M. Laugier, and J.M. Luck, Adv. Phys. {\bf 39}
(1990), 191.

[2] F. Brouers, J.P. Clerc, and G. Giraud, Phys. Rev. B {\bf 44} (1991), 5299;
F. Brouers, J.P. Clerc, G. Giraud, J.M. Laugier, and Z.A. Randriamanantany,
Phys. Rev. B {\bf 47} (1993), 666, and references therein.

[3] D.J. Bergman, Phys. Rep. {\bf 43} (1978), 377; Phys. Rev. Lett. {\bf 44}
(1980), 1285; Phys. Rev. B {\bf 23} (1981), 3058; Ann. Phys. {\bf 138} (1982),
78; G.W. Milton, Appl. Phys. Lett. {\bf 37} (1980), 300; J. Appl. Phys. {\bf
52} (1981), 5286; 5294; Phys. Rev. Lett. {\bf 46} (1981), 542.

[4] D.A.G. Bruggeman, Ann. Phys. (Leipzig) [Folge 5] {\bf 24} (1935), 636.

[5] R. Landauer, J. Appl. Phys. {\bf 23} (1952), 779.

[6] S. Kirkpatrick, Rev. Mod. Phys. {\bf 45} (1973), 574.

[7] I.M. Lifshitz, Adv. Phys. {\bf 13} (1964), 483; Sov. Phys. -- Uspekhi {\bf
7} (1965), 549. For a review, see: Th.M. Nieuwenhuizen, Physica {\bf 167 A}
(1990), 43.

[8] B. Hesselbo, D. Phil. Thesis (Oxford, 1994).

[9] F. Brouers, S. Blacher, and A.K. Sarychev, in {\it Fractals in the
Natural and Applied Sciences} (Kingston, 1995), p. 237.

[10] J.P. Clerc, G. Giraud, J.M. Laugier, J.M. Luck, and Z.A. Randriamanantany
(1991, unpublished).

[11] R.J. Wilson, {\it Introduction to Graph Theory} (Longman, London, 1979).

[12] R. Savit, Rev. Mod. Phys. {\bf 52} (1980), 453.

[13] D. Stauffer and A. Aharony, {\it Introduction to Percolation Theory}, 2nd
ed. (Taylor and Francis, London, 1992).

[14] J.L. Martin, in {\it Phase Transitions and Critical Phenomena}, vol. {\bf
3}, p. 97 (C. Domb and M.S. Green, eds.) (Academic Press, London, 1974).

[15] G. Parisi and N. Sourlas, Phys. Rev. Lett. {\bf 46} (1981), 871.

[16] T. Nagatani, J. Phys. C {\bf 14} (1981), 3383.

[17] M.H. Ernst, P.F.J. van Velthoven, and Th.M. Nieuwenhuizen, J. Phys. A {\bf
20} (1987), 949.

[18] Y.P. Pellegrini and J.M. Luck (1992, unpublished).

[19] J.M. Luck, Phys. Rev. B {\bf 43} (1991), 3933.

[20] J.P. Clerc, G. Giraud, J.M. Laugier, and J.M. Luck, J. Phys. A {\bf 18}
(1985), 2565.

[21] F. Brouers, D. Rauw, and J.P. Clerc, Physica {\bf 207 A} (1994), 249;
F. Brouers, D. Rauw, J.P. Clerc, and J.M. Laugier, Physica {\bf 207 A} (1994),
258.

[22] F. Spitzer, {\it Principles of Random Walk}, Graduate Texts in
Mathematics, vol. {\bf 34} (Springer, Berlin, 1976).

\vfill\eject
{\bf CAPTIONS OF FIGURES AND TABLES}
\bigskip
{\bf Figure 1:}
Schema of the sample used in this work.

{\bf Figure 2:}
Three examples of clusters (full lines),
together with their duals (dotted lines).

{\bf Figure 3:}
Plot of the resonance positions of the first 14 linear clusters.
The dashed line shows the upper band-edge $\lambda_L$ of eq. (3.26).

{\bf Figure 4:}
Histogram plots of the resonance spectra of all lattice animals
with (a) 7 bonds, (b) 9 bonds, and (c) 11 bonds.
The letters show the upper band-edges $\lambda_L$ and $\lambda_T$
of linear and $T$-shaped clusters, respectively.

{\bf Figure 5:}
Construction rules of the fractal clusters considered in section 3.6.

{\bf Figure 6:}
Largest generation whose resonance spectrum has been evaluated numerically,
for (a) the Worm, (b) the Cross, and (c) the Gasket.

{\bf Figure 7:}
Logarithmic plot of the small-$\lambda$ range of the resonance spectra of
the largest two generations of (a) the Worm, (b) the Cross, and (c) the Gasket.
The slopes of the full lines are the exact theoretical values of $\zeta$,
listed in Table 3.
\bigskip

{\bf Table 1:}
Positions of the resonances $\lambda_a$,
and associated cross-sections $\gamma_a$, of the clusters shown in Figure 2.

{\bf Table 2:}
Total numbers $N_n$ of lattice animals of $n$ bonds, up to $n_\max=11$.

{\bf Table 3:}
Scaling factors and exponents of the fractal clusters considered in
section 3.6.
}
\vfill\eject
\centerline {\bf Table 1}

$$\vbox{\init\halign to 14truecm
{\strut#&\vrule#\tabskip=1em plus 2em&
\hfil$#$\hfil&
\vrule#&
\hfil$#$\hfil&
\vrule#&
\hfil$#$\hfil&
\vrule#\tabskip 0pt\crr
&&\ &&\ &&\ &\cr
&&{\rm cluster}&&{\rm resonances}&&\hbox{cross-sections}&\cr
&&\ &&\ &&\ &\crr
&&\ &&\ &&\ &\cr
&&{\rm(a)}&&\matrix{
\lambda_1=0.23605\cr
\lambda_2=0.31285\cr
\lambda_3=0.59706\cr
\lambda_4=0.85404\cr
}&&\matrix{
\gamma_1=5.18968\cr
\gamma_2=2.17581\cr
\gamma_3=2.42350\cr
\gamma_4=0.10714\cr
}&\cr
&&\ &&\ &&\ &\crr
&&\ &&\ &&\ &\cr
&&{\rm(b)}&&\matrix{
\lambda_1=0.13067\cr
\lambda_2=0.30756\cr
\lambda_3=0.50000\cr
\lambda_4=0.69244\cr
\lambda_5=0.86933\cr
}&&\matrix{
\gamma_1=1.08684\cr
\gamma_2=0.48889\cr
\gamma_3=10.1794\cr
\gamma_4=0.48889\cr
\gamma_5=1.08684\cr
}&\cr
&&\ &&\ &&\ &\crr
&&\ &&\ &&\ &\cr
&&{\rm(c)}&&\matrix{
\lambda_1=0.24692\cr
\lambda_2=0.58281\cr
\lambda_3=0.64009\cr
\lambda_4=0.68901\cr
\lambda_5=0.84116\cr
}&&\matrix{
\gamma_1=12.9772\cr
\gamma_2=1.11399\cr
\gamma_3=5.17747\cr
\gamma_4=0.49606\cr
\gamma_5=0.12694\cr
}&\cr
&&\ &&\ &&\ &\crr
}}$$
\vfill\eject
\centerline {\bf Table 2}
$$\vbox{\init\halign to 8truecm
{\strut#&\vrule#\tabskip=1em plus 2em&
\hfil$#$\hfil&
\vrule#&
\hfil$#$\hfil&
\vrule#\tabskip 0pt\crr
&&\ &&\ &\cr
&&n &&N_n&\cr
&&\ &&\ &\crr
&&\ &&\ &\cr
&&\hfill 1 &&\hfill 2 &\cr
&&\hfill 2 &&\hfill 6 &\cr
&&\hfill 3 &&\hfill 22 &\cr
&&\hfill 4 &&\hfill 88 &\cr
&&\hfill 5 &&\hfill 372 &\cr
&&\hfill 6 &&\hfill 1\,628 &\cr
&&\hfill 7 &&\hfill 7\,312 &\cr
&&\hfill 8 &&\hfill 33\,466 &\cr
&&\hfill 9 &&\hfill 155\,446 &\cr
&& 10 &&\hfill 730\,534 &\cr
&& 11 &&\hfill 3\,466\,170 &\cr
&&\ &&\ &\crr
}}$$
\bigskip\null\bigskip
\centerline{\bf Table 3}
$$\vbox{\init\halign to 15truecm
{\strut#&\vrule#\tabskip =1em plus 2em&
\hfil$#$\hfil&
\vrule#&
\hfil$#$\hfil&
\vrule#&
\hfil$#$\hfil&
\vrule#&
\hfil$#$\hfil&
\vrule#&
\hfil$#$\hfil&
\vrule#&
\hfil$#$\hfil&
\vrule#\tabskip 0pt\crr
&&\ &&\ &&\ &&\ &&\ &&\ &\cr
&&{\rm Fractal}&&a&&b&&D_F&&t/\nu&&\zeta&\cr
&&\ &&\ &&\ &&\ &&\ &&\ &\crr
&&\ &&\ &&\ &&\ &&\ &&\ &\cr
&&{\rm Worm}&&3&&\sqrt{5}&&2\ln 3/\ln 5&&2\ln 3/\ln 5&&1&\cr
&&\ &&\ &&\ &&=1.36521&&=1.36521&&\ &\cr
&&\ &&\ &&\ &&\ &&\ &&\ &\crr
&&\ &&\ &&\ &&\ &&\ &&\ &\cr
&&{\rm Cross}&&5&&3&&\ln 5/\ln 3&&1&&\ln 3/\ln 5&\cr
&&\ &&\ &&\ &&=1.46497&&\ &&=0.68261&\cr
&&\ &&\ &&\ &&\ &&\ &&\ &\crr
&&\ &&\ &&\ &&\ &&\ &&\ &\cr
&&{\rm Gasket}&&3&&2&&\ln 3/\ln 2&&\ln(5/3)/\ln 2&&\ln(5/3)/\ln 3&\cr
&&\ &&\ &&\ &&=1.58496&&=0.73697&&=0.46497&\cr
&&\ &&\ &&\ &&\ &&\ &&\ &\crr
}}$$
\bye